\documentclass[twoside,twocolumn,9pt]{article}
\usepackage{extsizes}
\usepackage[maxauthors=8]{rsc}
\usepackage[version=3]{mhchem}
\usepackage[left=1.5cm, right=1.5cm, top=1.785cm, bottom=2.0cm]{geometry}
\usepackage{balance}
\usepackage{packages/widetext}
\usepackage{times,mathptmx}
\usepackage{sectsty}
\usepackage{graphicx}
\usepackage{lastpage}
\usepackage[format=plain,justification=raggedright,singlelinecheck=false,font={stretch=1.125,small,sf},labelfont=bf,labelsep=space]{caption}
\usepackage{float}
\usepackage{fancyhdr}
\usepackage{fnpos}
\usepackage[english]{babel}
\usepackage{array}
\usepackage{droidsans}
\usepackage{charter}
\usepackage[T1]{fontenc}
\usepackage[usenames,dvipsnames]{xcolor}
\usepackage{setspace}
\usepackage[compact]{titlesec}

\usepackage{url,subcaption,caption,nicefrac}

\definecolor{cream}{RGB}{222,217,201}

\begin{document}

\pagestyle{fancy}
\thispagestyle{plain}
\fancypagestyle{plain}{

\fancyhead[C]{\includegraphics[width=18.5cm]{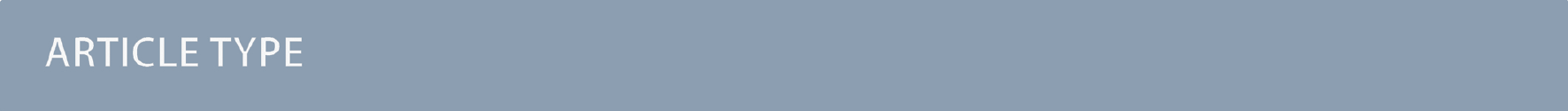}}
\fancyhead[L]{\hspace{0cm}\vspace{1.5cm}\includegraphics[height=30pt]{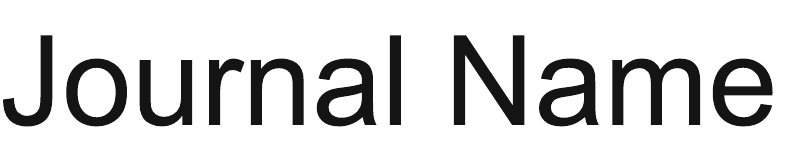}}
\fancyhead[R]{\hspace{0cm}\vspace{1.7cm}\includegraphics[height=55pt]{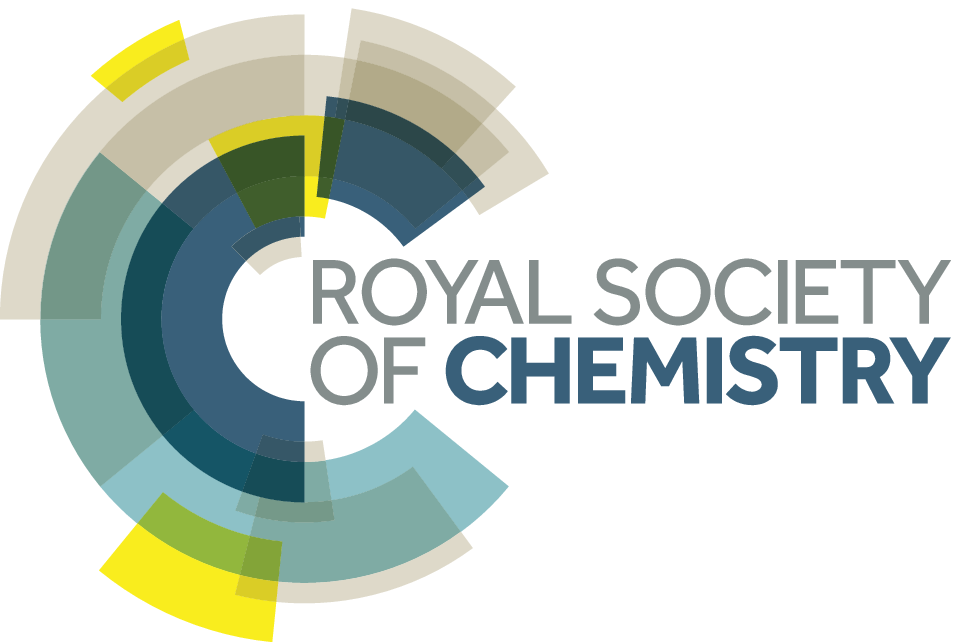}}
\renewcommand{\headrulewidth}{0pt}
}

\makeFNbottom
\makeatletter
\renewcommand\LARGE{\@setfontsize\LARGE{15pt}{17}}
\renewcommand\Large{\@setfontsize\Large{12pt}{14}}
\renewcommand\large{\@setfontsize\large{10pt}{12}}
\renewcommand\footnotesize{\@setfontsize\footnotesize{7pt}{10}}
\makeatother

\renewcommand{\thefootnote}{\fnsymbol{footnote}}
\renewcommand\footnoterule{\vspace*{1pt}%
\color{cream}\hrule width 3.5in height 0.4pt \color{black}\vspace*{5pt}}
\setcounter{secnumdepth}{5}

\makeatletter
\renewcommand\@biblabel[1]{#1}
\renewcommand\@makefntext[1]%
{\noindent\makebox[0pt][r]{\@thefnmark\,}#1}
\makeatother
\renewcommand{\figurename}{\small{Fig.}~}
\sectionfont{\sffamily\Large}
\subsectionfont{\normalsize}
\subsubsectionfont{\bf}
\setstretch{1.125} 
\setlength{\skip\footins}{0.8cm}
\setlength{\footnotesep}{0.25cm}
\setlength{\jot}{10pt}
\titlespacing*{\section}{0pt}{4pt}{4pt}
\titlespacing*{\subsection}{0pt}{15pt}{1pt}

\fancyfoot{}
\fancyfoot[LO,RE]{\vspace{-7.1pt}\includegraphics[height=9pt]{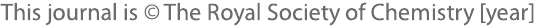}}
\fancyfoot[CO]{\vspace{-7.1pt}\hspace{13.2cm}\includegraphics{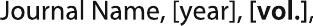}}
\fancyfoot[CE]{\vspace{-7.2pt}\hspace{-14.2cm}\includegraphics{head_foot/RF}}
\fancyfoot[RO]{\footnotesize{\sffamily{1--\pageref{LastPage} ~\textbar  \hspace{2pt}\thepage}}}
\fancyfoot[LE]{\footnotesize{\sffamily{\thepage~\textbar\hspace{3.45cm} 1--\pageref{LastPage}}}}
\fancyhead{}
\renewcommand{\headrulewidth}{0pt}
\renewcommand{\footrulewidth}{0pt}
\setlength{\arrayrulewidth}{1pt}
\setlength{\columnsep}{6.5mm}
\setlength\bibsep{1pt}

\makeatletter
\newlength{\figrulesep}
\setlength{\figrulesep}{0.5\textfloatsep}

\newcommand{\topfigrule}{\vspace*{-1pt} \noindent{\color{cream}\rule[-\figrulesep]{\columnwidth}{1.5pt}} }
\newcommand{\botfigrule}{\vspace*{-2pt} \noindent{\color{cream}\rule[\figrulesep]{\columnwidth}{1.5pt}} }
\newcommand{\dblfigrule}{\vspace*{-1pt} \noindent{\color{cream}\rule[-\figrulesep]{\textwidth}{1.5pt}} }

\makeatother

\twocolumn[
\begin{@twocolumnfalse}
\vspace{3cm}
\sffamily
\begin{tabular}{m{4.5cm} p{13.5cm} }

\includegraphics{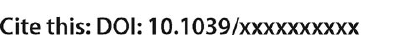} & \noindent\LARGE{\textbf{ Density Functional Based Simulations of Proton Permeation of Graphene and Hexagonal Boron Nitride }} \\
\vspace{0.3cm} & \vspace{0.3cm} \\

& \noindent\large{J.M.H.~Kroes$^{\ast}$, A.~Fasolino, and M.I.~Katsnelson} \\

\includegraphics{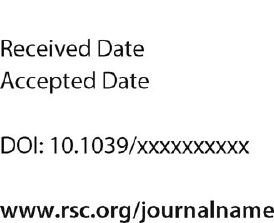} & \noindent\normalsize{ \textbf
{
Using density functional theory, we study proton permeation through graphene and hexagonal boron nitride.
We consider several factors influencing the barriers for permeation, including structural optimization, the role of the solvent, surface curvature and proton transport through hydrogenated samples.
Furthermore, we discuss the ground state charge transfer from the membrane to the proton and the strong tendency for bond formation.
If the process is assumed to be slow we find that none of these effects lead to a satisfactory answer to the observed discrepancies between theory and experiment.
}} \\

\end{tabular}

\end{@twocolumnfalse} \vspace{0.6cm}

]

\renewcommand*\rmdefault{bch}\normalfont\upshape
\rmfamily
\section*{}
\vspace{-1cm}


\footnotetext{\textit{Radboud University, Institute for Molecules and Materials, Heijendaalseweg 135, 6525~AJ Nijmegen, The Netherlands.}}



\section{Introduction}

\begin{figure}[bht]\centering\small \hfill
\begin{subfigure}{0.18\textwidth} \includegraphics[width=\textwidth]{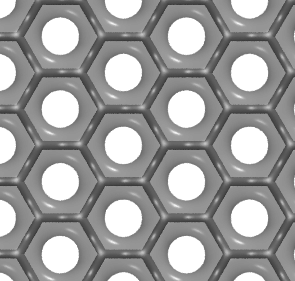} \caption{\centering graphene}\end{subfigure} \hfill
\begin{subfigure}{0.18\textwidth} \includegraphics[width=\textwidth]{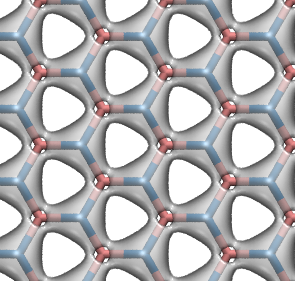} \caption{\centering h-BN}\end{subfigure} \hfill
\hfill
\caption{Electron density clouds at an isovalue of 0.1~e$^-$/bohr$^3$ for monolayer graphene (a) and h-BN (b) with B (N) in pink (blue) . \label{fig:clouds}}
\end{figure}

Graphene and hexagonal boron nitride (h-BN) are chemically inert and strongly hydrophobic materials.
As a result, it has generally been expected that no ions could penetrate their dense electron clouds (Fig.~\ref{fig:clouds}).
However, recently it was nevertheless observed~\cite{hu_proton_2014} that it is possible for protons to do precisely that.

In the experiments by Hu et al.~\cite{hu_proton_2014}, hydrogen was split into electrons and protons, with protons passing through the membrane and electrons passing through an electrical wire under the influence of an applied bias voltage. 
These protons and electrons then recombine into hydrogen again on the other side. 
From the resulting IV-curve the conductivity could then be extracted. 
The temperature dependence of the conductivity was used to determine the energy barrier for permeation. 
These experiments are characterized by a highly reproducible and linear IV-behaviour,
indicative of an independence of specific details of the sample such as defects. 

Hu et al. found proton permeation to be easiest for monolayer h-BN but also observable for bilayer h-BN and monolayer graphene~\cite{hu_proton_2014}.
Notably, no measurable conductance was observed for bilayer graphene, which was understood to be a result of its AB layered stacking. 
From the near-perfect Boltzmann behaviour of the conductance as a function of temperature the associated barriers were estimated to be 0.3, 0.6 and 0.8~eV for monolayer h-BN, bilayer h-BN and monolayer graphene respectively.
At a typical bias voltage of 0.1~V, the resulting proton current through h-BN corresponds to roughly 600 protons/s/nm$^2$.
We therefore consider this process to be slow in comparison to typical ionic timescales and attempt rates.

More recently it has also been observed that the permeation barrier depends on the ionic mass~\cite{lozada-hidalgo_sieving_2016}, making these materials naturally selective membranes for separating protons and deuterons. 
This was explained~\cite{lozada-hidalgo_sieving_2016} as originating from the isotope effect, i.e. the difference in zero-point energies of the proton and deuteron.
The zero-point energy, $\Delta E^{0}$, comes into play only for the initial state, where the proton is bound to an oxygen atom. In water, $\Delta E^{0}$ is principally determined by the OH$^-$ stretching frequency, which is estimated to be $\Delta E^{0}=0.20$~eV for protons and $\Delta E^{0}=0.14$~eV for deuterons based on literature values~\cite{wiberg_deuterium_1955,lozada-hidalgo_sieving_2016}.
In DFT, the barrier ($\Delta E$) is calculated as difference between the transition state and the initial state (without zero-point energy), whereas experimentally, the barrier is calculated between the transition state and the initial state which is raised by its zero-point energy with respect to the value given by DFT. Therefore, the barrier calculated within DFT is larger than the experimental barrier which naturally includes the zero-point motion, namely
$\Delta E^\textrm{exp} = \Delta E^\textrm{DFT} - \Delta E^0.$

This selective permeation of protons versus other ions as well as the separation of protons and deuterons, has multiple direct technological applications making a fundamental understanding of this process important (see also Refs.~\cite{hu_proton_2014,lozada-hidalgo_sieving_2016} and references therein). 
The most obvious application lies perhaps in proton exchange membrane fuel cells, which can be scaled down in size by the use of such nano-sized membranes. 
Another application may be found in the separation of protons and deuterons for example in the production of heavy water. 
This is typically a complicated and costly process which may be greatly simplified by the use of naturally semi-permeable membranes.

On the theoretical side, several efforts have been undertaken to understand the proton permeation of 2D membranes~\cite{hu_proton_2014,wang_graphene_2010,miao_first_2013,tsetseris_graphene:_2014}, most of them using density functional theory (DFT). 
However, significant discrepancies remain between the barriers found in experiment and those calculated. 
For example, for graphene, calculated values are 
  1.17~eV~\cite{wang_graphene_2010}, 
  1.41 or 2.21~eV~\cite{miao_first_2013} (the latter after relaxation),
  1.25--1.40~eV~\cite{hu_proton_2014} or
  1.56~eV~\cite{seel_proton_2016}, all of which overestimate the experimental value of 1.0~eV~\cite{lozada-hidalgo_sieving_2016}.
  More severe discrepancies arise when attempting to improve upon these calculations as discussed in the following sections.

One may question the accuracy of the applied theoretical methods for the problem at hand and whether DFT based calculations are indeed suitable to describe the permeation of a proton through a highly polarizable membrane or the quantum nature of the ions or excited states should be taken into account. 
Classical molecular dynamics simulations~\cite{despiau-pujo_elementary_2013} using the REBO potential~\cite{brenner_second-generation_2002} found much higher barriers for hydrogen permeation (11 eV for REBO compared to 2.5 eV for DFT) illustrating the need to go beyond classical force fields for quantitative estimates of the barriers. 
Recent calculations also aimed to take into account the quantum nature of the ions by means of path integral molecular dynamics~\cite{poltavsky_quantum_2016}. These calculations predict lower barriers as result of quantum tunneling but at the same time predict a much larger isotope effects ($\sim$1~eV instead of 0.06~eV) than observed experimentally. Therefore, despite the claims made in Ref.~\cite{poltavsky_quantum_2016}, the puzzle of low barriers is still unsolved. 

Aside from these theoretical considerations, many of the details of the experimental setup may prove essential. 
The theoretical difficulty to reproduce these barriers stems at least partially from the difficulty in identifying the relevant experimental parameters. 
For example, the role of the solvent (Nafion) in which the proton diffuses is largely unknown.
The effect of temperature, surface curvature or applied electric field may also prove crucial.
Finally, it can also not be excluded that (ad-)atoms or other defects are present that may modify the barrier as suggested by experiments with varying pH concentrations~\cite{achtyl_aqueous_2015,walker_measuring_2015}.

\section{Methods}

Here we consider various factors relevant to the calculated proton permeation barriers within the framework of DFT.
Our calculations were done using the non-empirical PBE exchange-correlation functional~\cite{perdew_generalized_1996,perdew_generalized_1997} implemented in the CP2K~code~\cite{vandevondele_quickstep:_2005}.
To account for van-der-Waals interactions, especially relevant for the simulations of the membrane in water, we use the Grimme-D3 dispersion correction term~\cite{grimme_consistent_2010}.
Unless mentioned, for graphene (h-BN) we use a model consisting of 6x12 orthorhombic cells, corresponding to 288C (144B and 144N) atoms and a supercell of 25.6$\times$29.5~\AA$^2$ (26.2$\times$30.3~\AA$^2$) with periodic boundary conditions. For the perpendicular cell size we use 15~\AA.
The lattice constant of graphene and h-BN are fixed such that $r_\textrm{CC} = 1.421$~\AA\ and $r_\textrm{BN}=1.457$~\AA\ when flat.
We considered non-spin polarized calculations at the $\Gamma$-point of the Brioullin zone , used a Fermi-Dirac electronic smearing with a width of 300~K. 
The Quickstep method is employed, with wave functions expanded onto a localized double-$\zeta$-valence-polarized basis set and the electronic density expanded onto a plane-wave basis set with a kinetic energy cutoff of 500 Ry. Goedecker-Teter-Hutter pseudopotentials~\cite{goedecker_separable_1996} are used to describe the interactions with the core electrons. 

For the optimization of the barrier we used climbing-image nudged elastic band (CI-NEB~\cite{henkelman_climbing_2000}) method with 18 images, of which one was the climbing image.

We also performed Born-Opperheimer molecular dynamics (MD) simulations of a proton, in water solution, permeating through graphene or h-BN.
For this we used a velocity-Verlet integration timestep of 0.5~fs.
These simulations were done using using a periodic model of graphene or h-BN consisting of 60~atoms with a solvent consisting of 88 water molecules in an orthorhombic super cell, corresponding to a 1~kg/L density of water when taking into account a 3.5~\AA\ vacuum due to the hydrophobic nature of the membrane.
The cell sizes are 12.8$\times$12.3$\times$20.2~\AA$^3$ for graphene and 13.2$\times$12.7$\times$19.4~\AA$^3$ for h-BN.
In order to have well converged wavefunctions, a 1~ps NVE benchmark simulation was performed to determine the necessary convergence criteria to have total energy drifts of at most 1~meV/ps, which was found to be $\epsilon^\textrm{SCF} = 10^{-7}$~Ha.
For temperature control we used the coloured noise NVT thermostat~\cite{bussi_canonical_2007} with a coupling constant of 1~ps and a constant temperature of 325~K.
After an equilibration period of 30~ps, a proton is pulled through the membrane by means of a steered molecular dynamics (SMD) simulation in 2~ps. 

To compute the free energy barrier at room temperature with SMD, a guiding spring-force is added between the proton and a (moving) target point using a collective variable, $\xi$, equal to the height difference between the proton and an atom in the membrane
\[ F^\mathrm{guide} = -\kappa\left[\xi^\mathrm{pr}(t)-\xi^\mathrm{pr}(0) - v_\xi t\right]^2,\]
with spring constant $\kappa = 10$~eV/\AA, where $v_\xi = 5$~\AA/ps is the target speed, chosen such that during the simulation time the proton will pass through the membrane.
The work done then gives an estimate of the barrier and is computed as 
\[ W(\xi) = \kappa v \int_0^t dt' \left[ \xi - \xi^\mathrm{target}(t') \right] dt'. \]
These simulations were done using the PLUMED package~\cite{bonomi_plumed:_2009}, integrated with CP2K.

\section{Static energy barriers}

\begin{figure*}[!th]\centering\small
\begin{subfigure}{0.63\textwidth} \includegraphics[height=39mm]{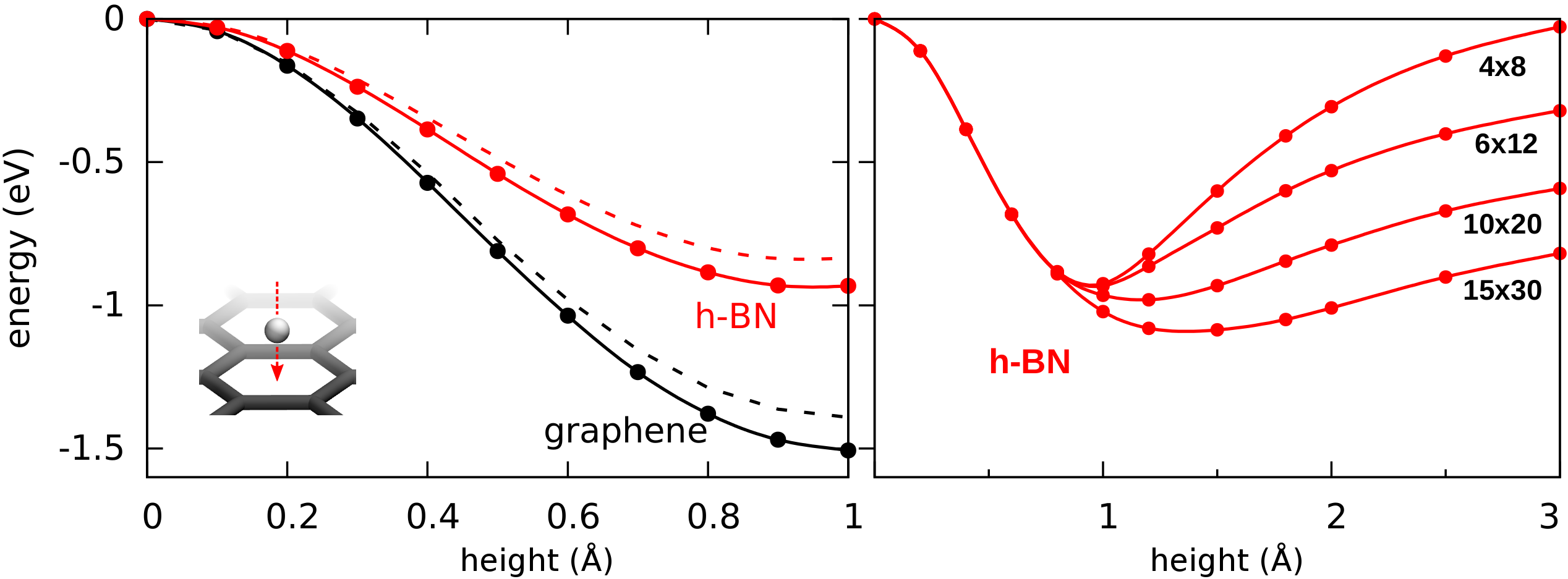} \caption{\centering static barriers}\end{subfigure} \hfill
\begin{subfigure}{0.36\textwidth} \includegraphics[height=39mm]{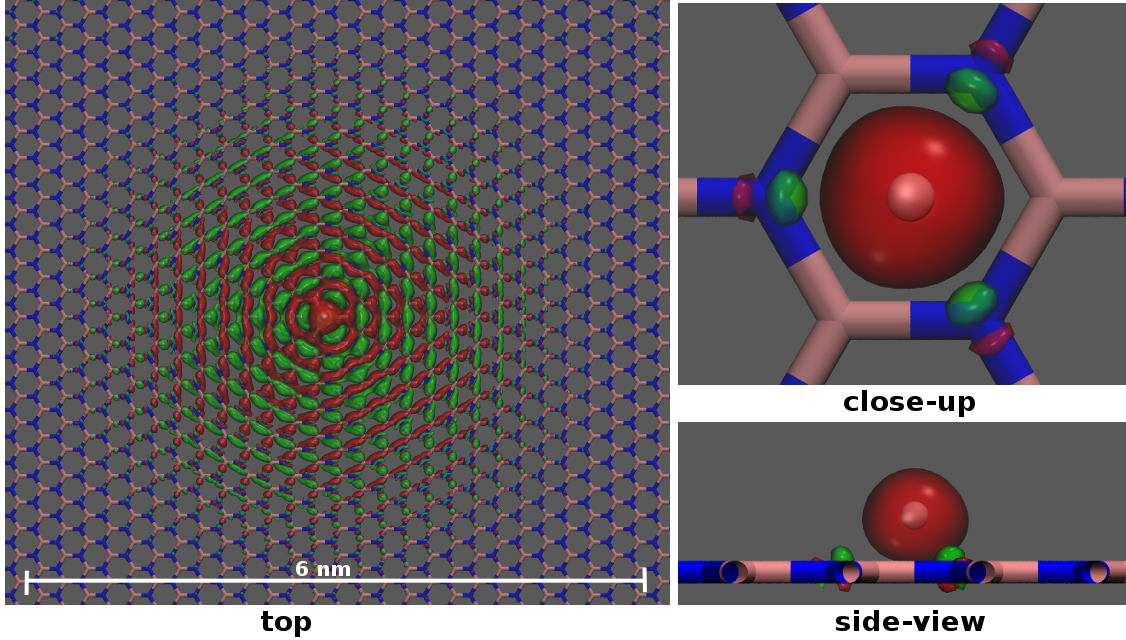}   \caption{\centering $\Delta\rho$ for h-BN with a proton at $h$=1~\AA}\end{subfigure}
\caption{(a) shows the static energy barrier of a proton passing through the center of a hexagon for h-BN (red) and graphene (black) without structural relaxation at equilibrium lattice constant (solid) and under 2\%\ isotropic tensile strain (dashed). In the left panel, also the Bader charge is shown for each frame (which assigns about half an electron to the proton) by the triangles connected by dotted lines.
The panel on the right shows the dependence on the model size for distances $h\ge$ 1~\AA\ is shown. 
(b) shows $\Delta\rho=\rho_\textrm{tot}-\rho_\textrm{h-BN}$ for (15$\times$30) h-BN  with a proton at $h$=1~\AA\ plotted at an isovalue of +/-~0.0001~e$^-$/bohr$^3$ (top-view) and +/-~0.01~e$^-$/bohr (side-views) in red/green. \label{static}}
\end{figure*}

\begin{figure}\centering\small
\begin{subfigure}{.14\textwidth} \includegraphics[height=16mm]{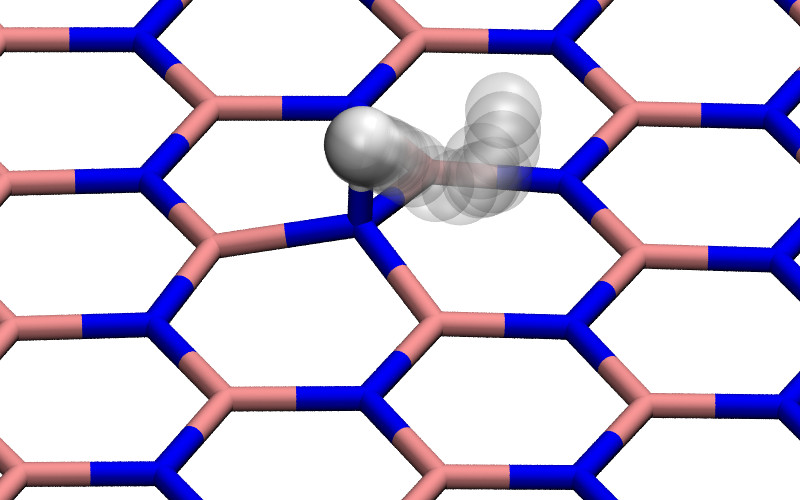} \caption{\centering optimization} \end{subfigure}\hfill
\begin{subfigure}{.14\textwidth} \includegraphics[height=16mm]{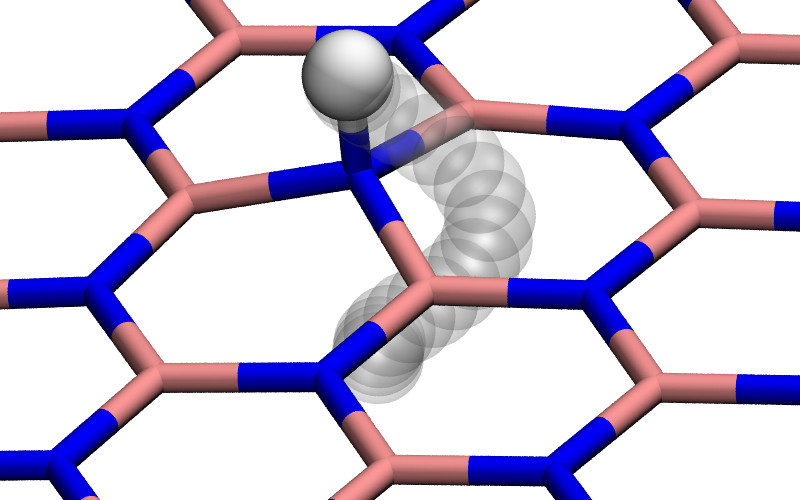} \caption{\centering NEB path} \end{subfigure}\hfill
\begin{subfigure}{.19\textwidth} \includegraphics[height=16mm]{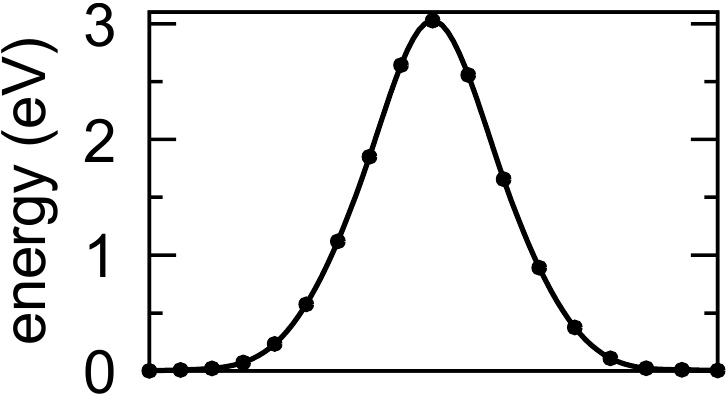} \caption{\centering NEB barrier} \end{subfigure}
\caption{\label{opt} In (a) the optimization of the initial state is illustrated. The proton goes from 3~\AA\ above a hexagon center to a state where it is bonded to boron with a bond length of 1.05~\AA. The system gains 4.6 eV during the optimization.
(b) shows the CI-NEB path for permeation between the two equivalent sides of the membranes. 
Boron, nitrogen and hydrogen atoms are shown in pink, blue and white respectively.
The energy barrier corresponding to (b) is shown in (c).
}
\end{figure}

In the simplest calculations~\cite{hu_proton_2014,wang_graphene_2010,miao_first_2013,tsetseris_graphene:_2014}, the proton passes through the center of a hexagon with the lattice remaining fixed and the proton permeation path being a straight line perpendicular to the material. 
The energy barriers found in this way are about 0.9 or 1.5 eV for a proton passing through h-BN or graphene respectively as shown in Fig.~\ref{static}a.
These values are higher than the experimentally observed barriers (0.5 and 1.0~eV), but appear to be in reasonable agreement. 
Tensile strain may further reduce the calculated barriers slightly but, as shown in Fig.~\ref{static}a, even 2\%\ of tensile strain reduces the calculated barriers by only about 0.1~eV. We verified that the perpendicular cell size is sufficient, increasing it from 15 to 30~\AA, the barrier changes less than 2~meV.

The interpretation of the proton by itself in DFT calculations is however not obvious.
This is because, in the presence of a highly polarizable membrane such as graphene or hBN, an electron will be taken from the membrane leaving a charged membrane with neutral hydrogen in the ground state as illustrated in Fig.~\ref{static}b and described in more detail in the supplementary information (Fig. S1, S2 and Table S1). As a result, at larger proton-membrane separation distances (above $\sim$1~\AA), the energy curves depend strongly on the model size as shown in the right panel of Fig.~\ref{static}a but flatten out with increasing sample size.

If we indeed assume this process to be slow, relaxation of the nuclear coordinates should be possible.
When such a relaxation is done using a quasi-Newton method~\cite{byrd_limited_1995}, the initial state is found to be unstable and the proton will chemisorb on the surface. If we then compute the barrier by means of the nudged elastic band (NEB) method with one climbing image (CI-NEB~\cite{henkelman_climbing_2000}) the barrier (for h-BN) increases from 0.5 to 3.0~eV as shown in Fig.~\ref{opt}, a value similar to the barrier for permeation of a hydrogen atom~\cite{kroes_energetics_2016}.
Thus, contrary to the intuition that optimization of the path should lead to a lower barrier, the barrier increases as a result of relaxation because the initial state is in fact unstable.

\section{Effect of the solvent}

\begin{figure*}\centering\small\hfill
\begin{subfigure}{0.14\textwidth} \centering \includegraphics[height=43mm]{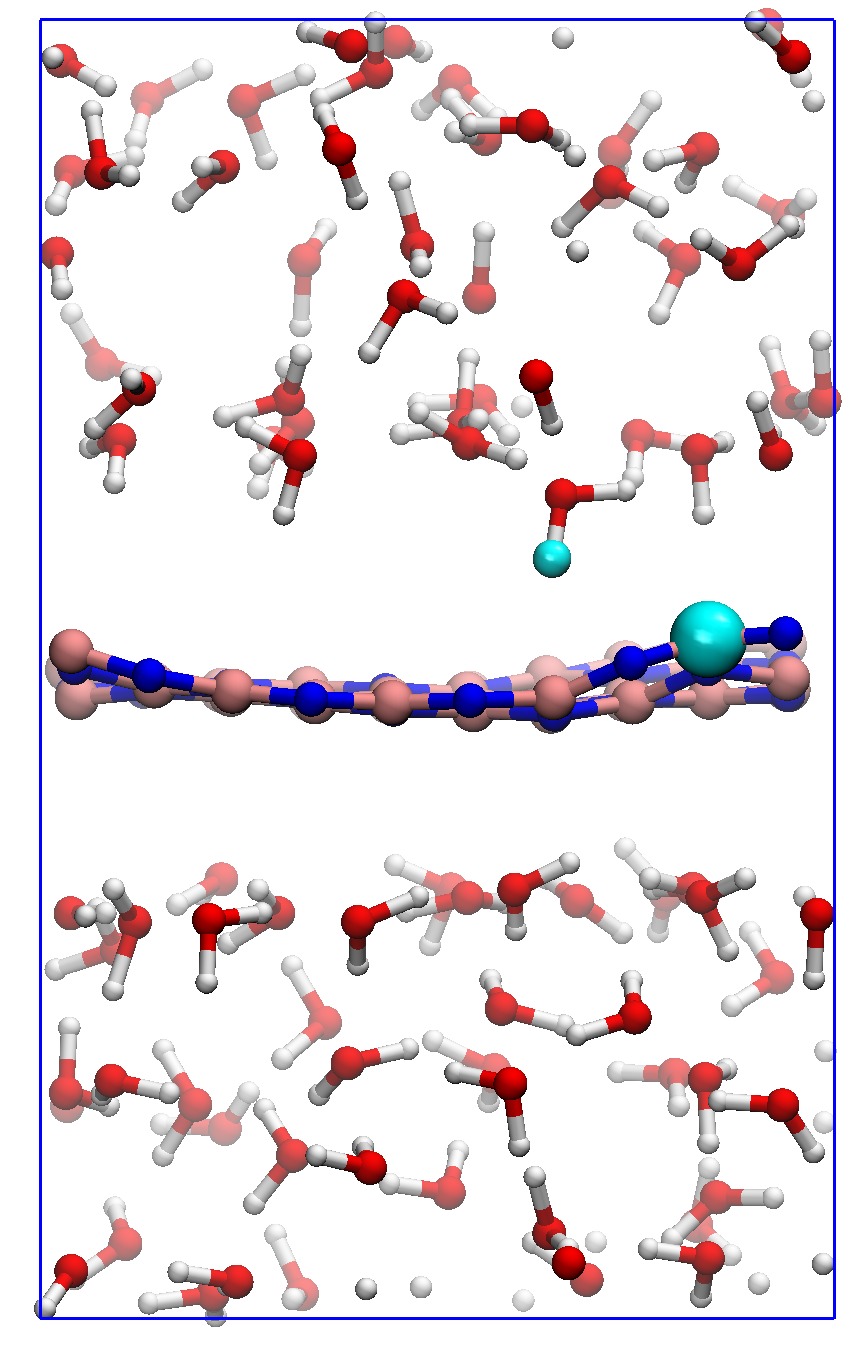} \caption{\centering $t_0$}\end{subfigure} \hfill
\begin{subfigure}{0.14\textwidth} \centering \includegraphics[height=43mm]{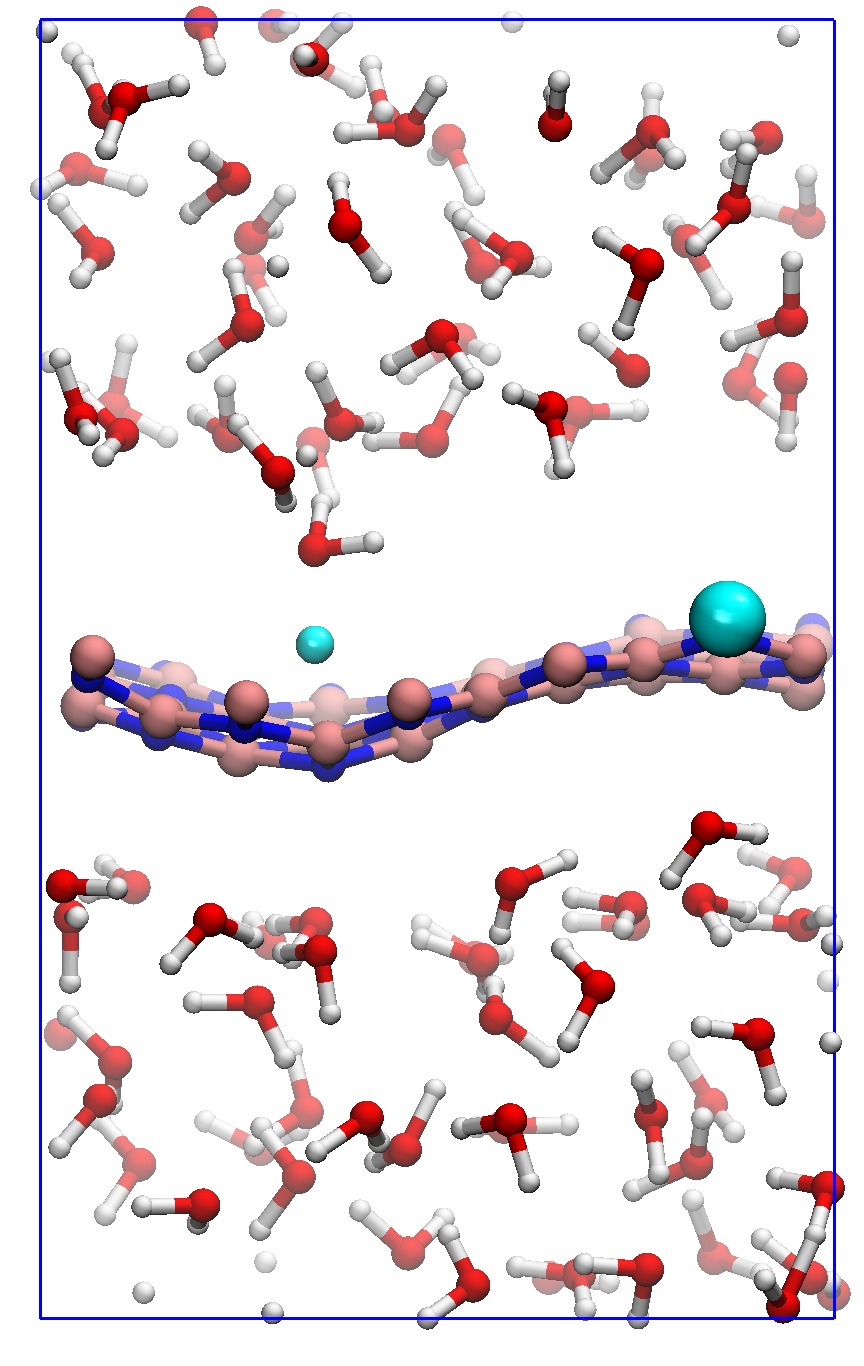} \caption{\centering $t_0$+0.4~ps}\end{subfigure} \hfill
\begin{subfigure}{0.14\textwidth} \centering \includegraphics[height=43mm]{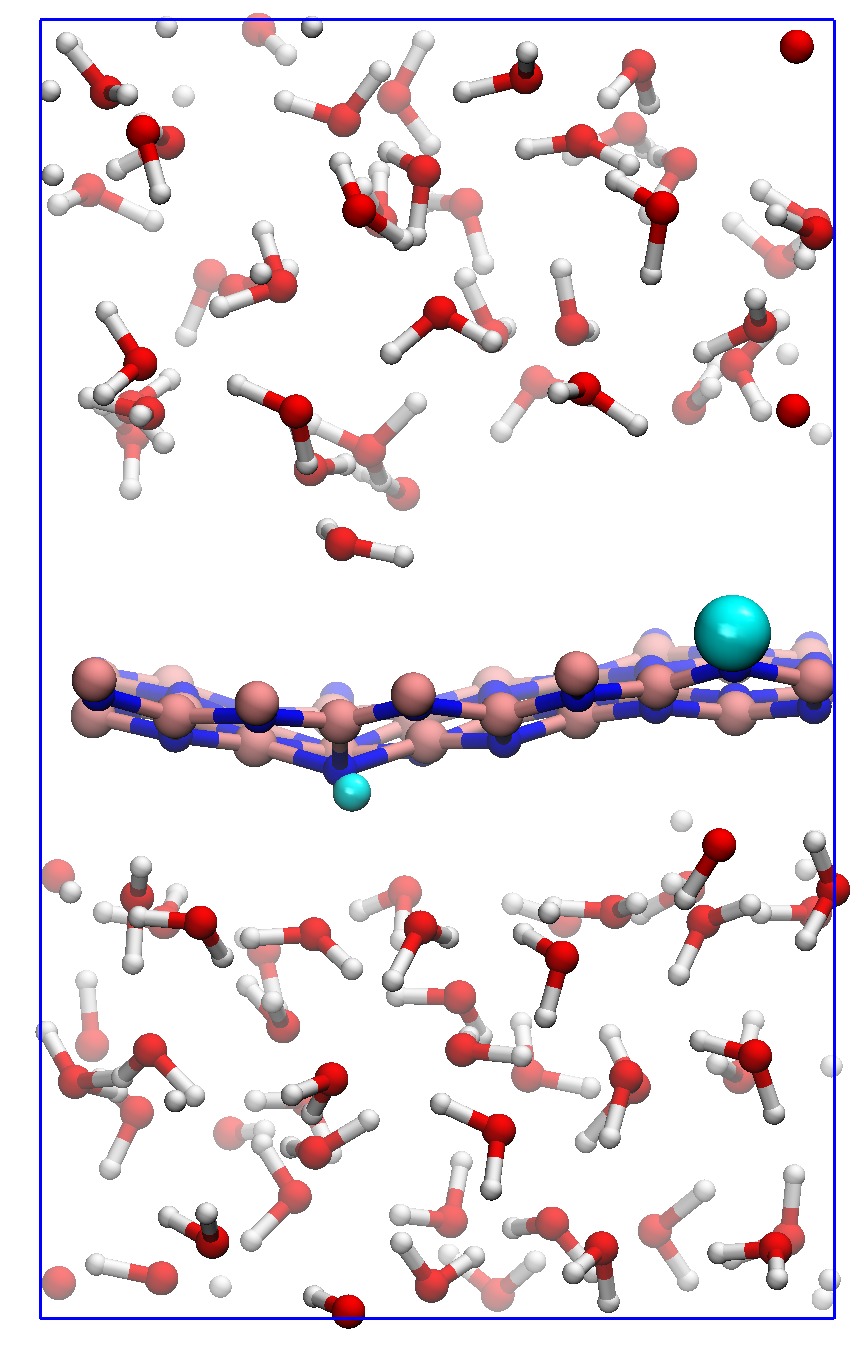} \caption{\centering $t_0$+0.8~ps}\end{subfigure} \hfill
\begin{subfigure}{0.14\textwidth} \centering \includegraphics[height=43mm]{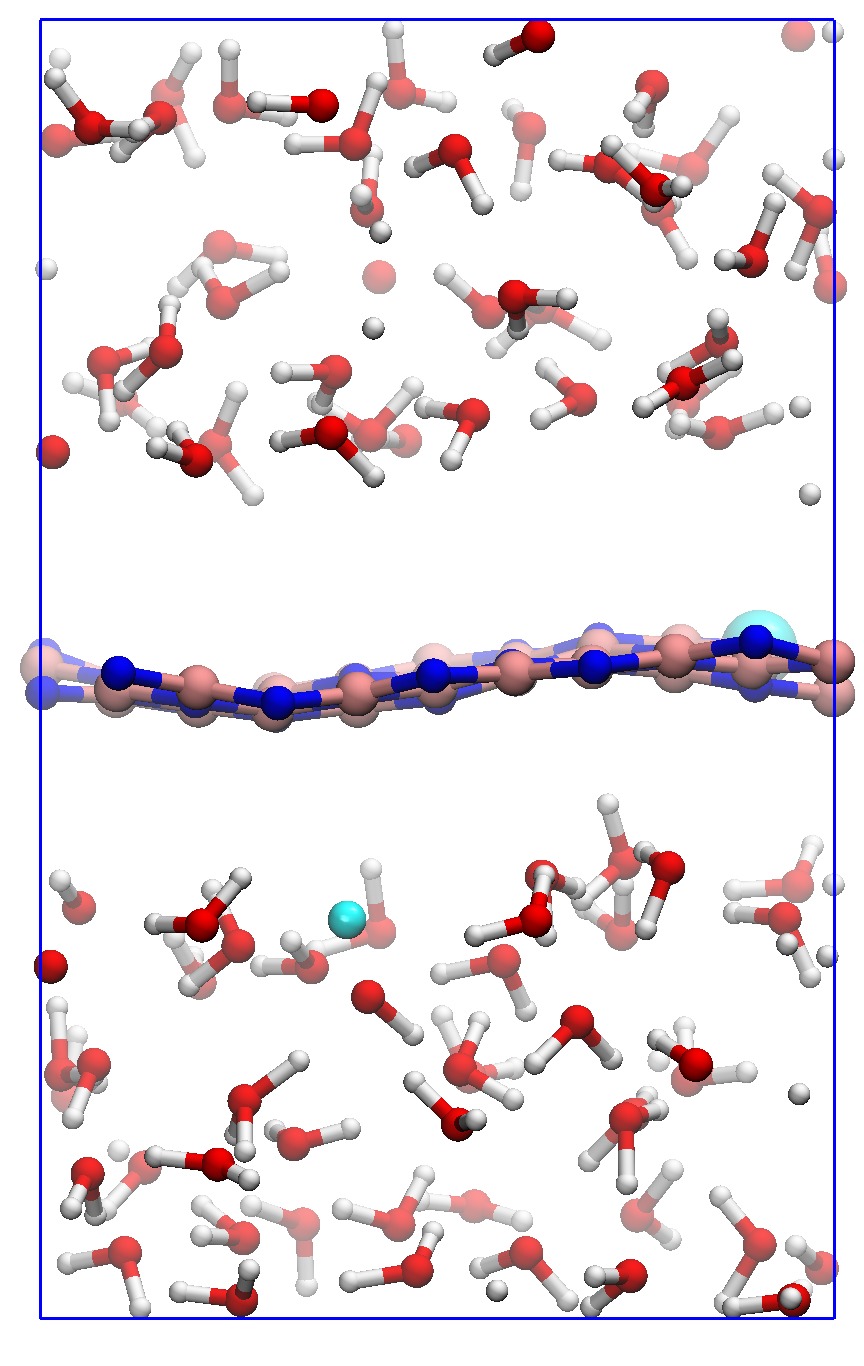} \caption{\centering $t_0$+1.2~ps}\end{subfigure} \hfill
\begin{subfigure}{0.14\textwidth} \centering \includegraphics[height=43mm]{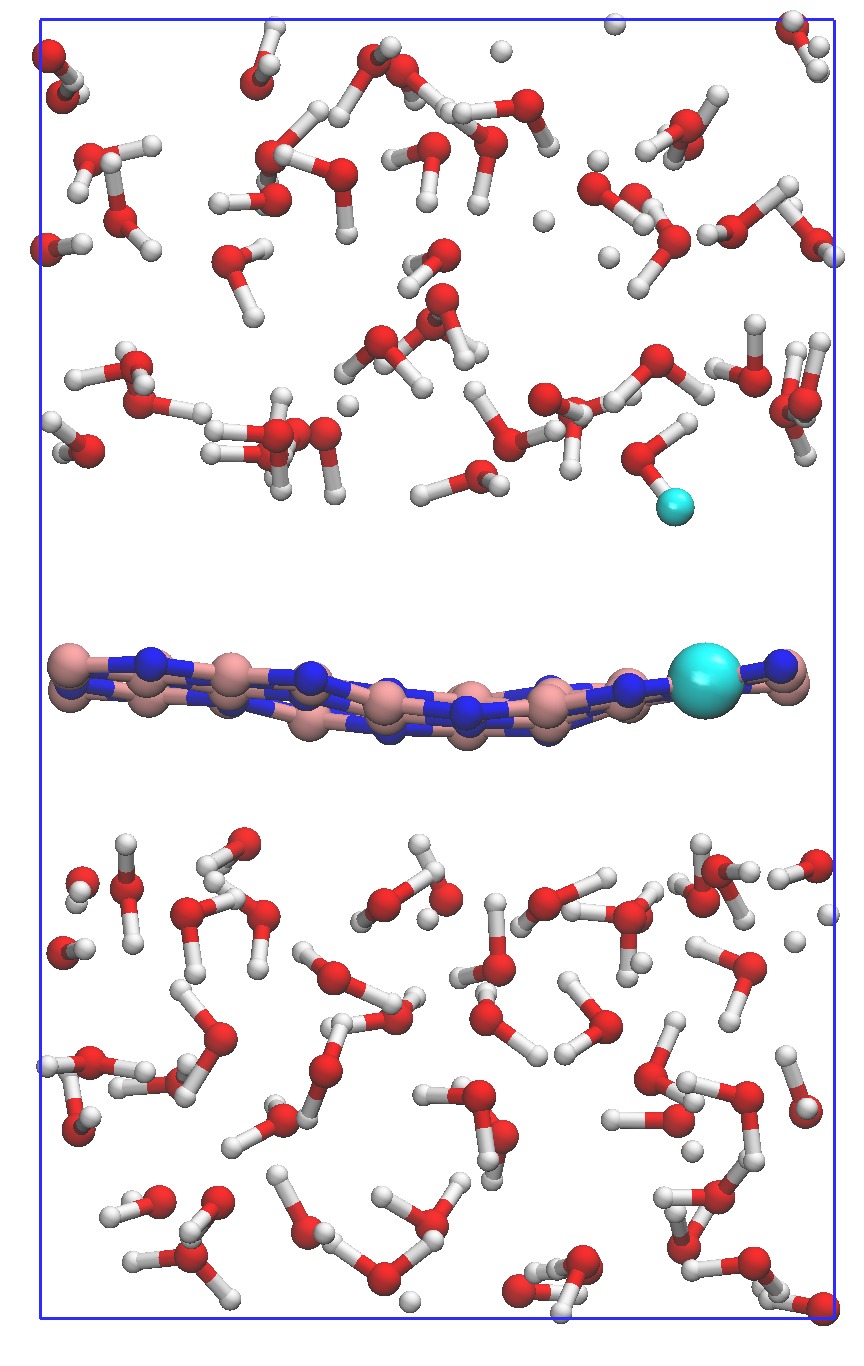} \caption{\centering $t_0$+1.6~ps}\end{subfigure} 
\begin{subfigure}{0.27\textwidth} \centering \includegraphics[height=43mm]{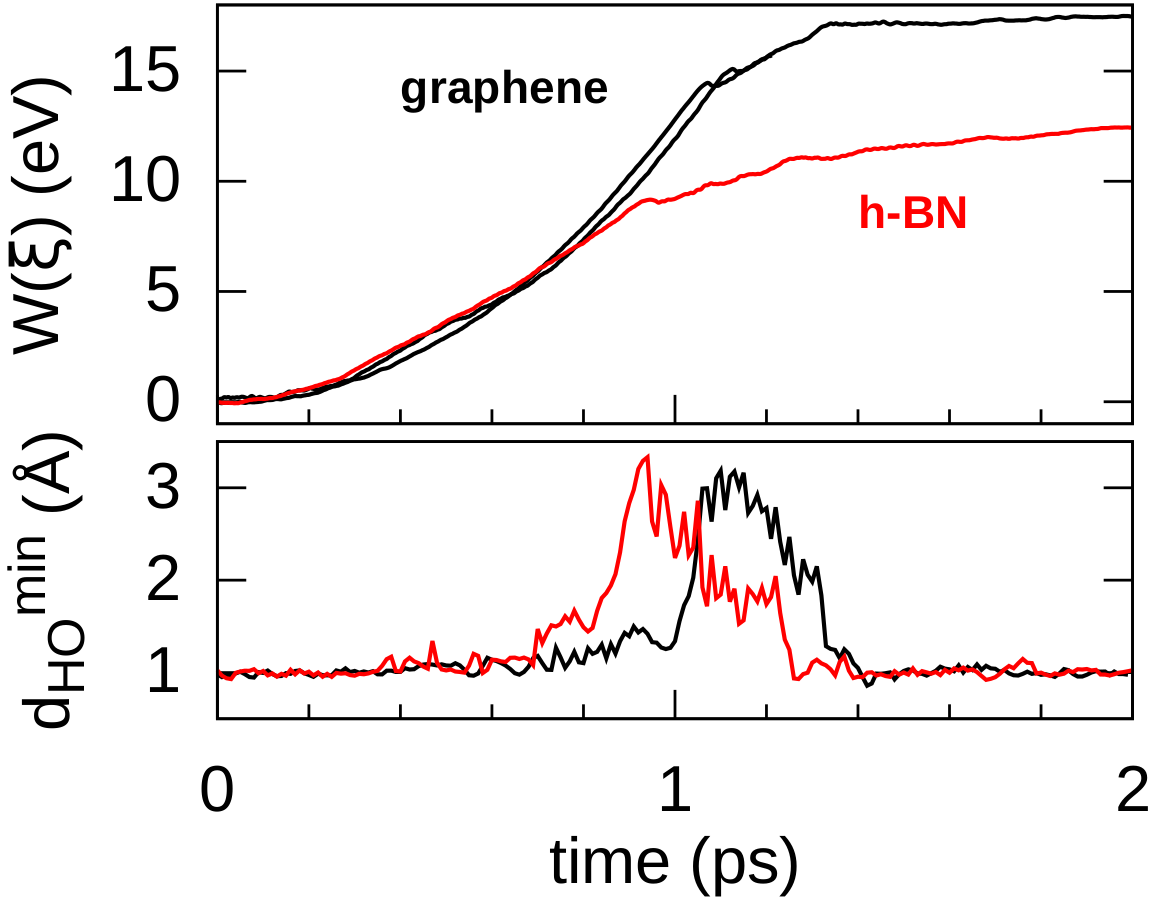}                              \caption{\centering W($\xi$)\label{W_xi}}\end{subfigure} \hfill
\hfill
\caption{Time lapse (a-e) of the SMD simulation to force proton permeation for h-BN.
B, N, O and H atoms are shown in pink, blue, red and white respectively.
Light-blue balls mark the atoms to which the SMD force is applied.
(f) shows the work done by the SMD force and the distance from the forced nucleus to the nearest oxygen.
In red (black) the results for h-BN (graphene). 
The two black lines indicate different H-C pairs for permeation.
\label{SMD-frames}}
\end{figure*}

In order to consider the role of the solvent we performed room-temperature Born-Oppenheimer \emph{ab initio} molecular dynamics simulations of the membrane plus the explicit solvent with one additional solved proton.
An example of the proton permeation process in the presence of water for h-BN is shown in Fig.~\ref{SMD-frames}.
In this case we consider proton-membrane separation distances larger than 1~\AA\ because the proton is stabilized by the presence of the water as demonstrated by the Mulliken population analysis~\cite{mulliken_electronic_1955} in Table S2. 

After equilibration, the hydrogen nucleus closest to the membrane, together with the nearest atom in the lattice (B or N) are chosen for the SMD forcing. 
The SMD spring force is applied to both these atoms, forcing a change in z-coordinate difference (perpendicular to the membrane) from its initial value to a value of opposite sign and large enough that the SMD work is expected to flatten out. 
We note that, because the excess charge is spread over different nuclei in the liquid (see Fig. S3 and Table S2), the identification of a single proton is not possible and our selected H nucleus is in fact close to neutral initially. 
Rather than passing directly through the membrane, we observe the selected atom to move sideways away from its initial position. Simultaneously, a curvature is induced in the membrane (Fig. 4b) which brings the atoms closer in height to follow the (moving) SMD equilibrium spring position. This shows that it is favourable to bend the substrate rather than to directly approach the surface. 
The resulting work done is shown in Fig.~\ref{W_xi}. 
The barrier estimated from these curves contains two separate effects. The first involves the removal of the proton from the liquid and the second involves passing through the membrane.
These steered dynamics simulations are equivalent to the Tomlinson model (see e.g. Ref.~\cite{Sasaki_tomlinson_1996} and references therein) used in the field of friction at the atomic scale. The top of the barrier relevant to the experimental setup is therefore defined by the moment of slipping, whereas the starting configuration is given by the moment of OH bond breaking. As can be seen from Fig.\ref{W_xi} and in more detail in Fig. S4, these barriers are then of the order of 3--5 eV, in qualitative agreement the nudged elastic band calculations of Ref.~\cite{achtyl_aqueous_2015}.
After the reformation of an OH bond on the opposite side of the membrane the SMD work done indeed tends to flatten out as expected.
Compared to the static calculations, rather than decreasing the barrier, barriers are thus increased because of the stabilizing effect the water has on the proton, lowering the energy of the initial state.
We thus conclude that the inclusion of water in our model (one proton for 87 water molecules, corresponding to a pH close to zero) cannot explain the observed discrepancy between theory and experiment.

\section{Curvature}

\begin{figure} \centering\small
\includegraphics[width=.40\textwidth]{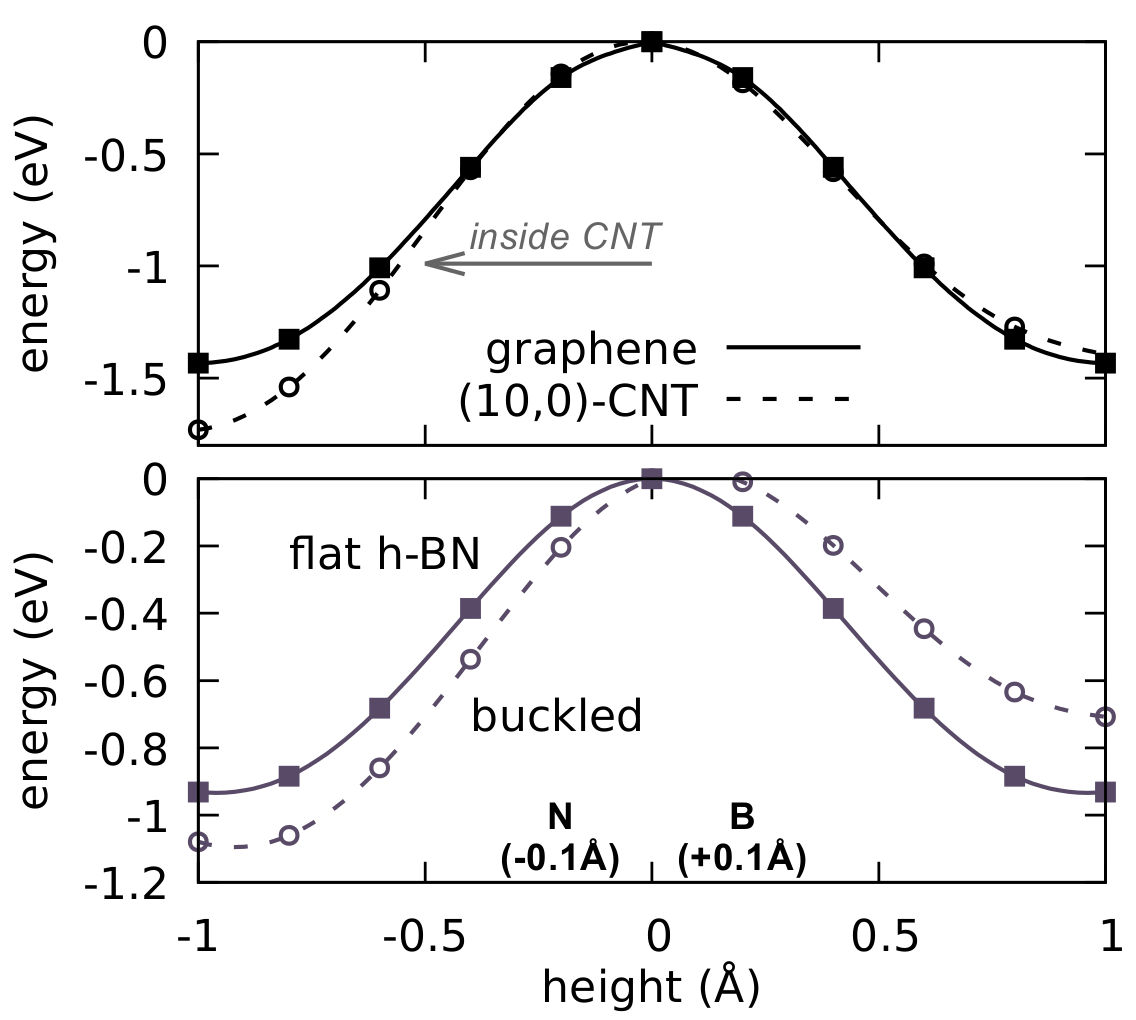}
\caption{Energy barrier for proton permeation for the indicated curvature models (see text).\label{curv}}
\end{figure}

We also considered the effect of curvature.
For this we return to the case of static permeation, i.e. without optimization of the path.
We consider two different cases of curvature, namely (i) a (10,0)--carbon nanotube (CNT) compared to graphene, and (ii) a buckled h-BN surface, where all B (N) atoms are displaced up (down) by 0.1~\AA. The latter configuration could for example be the result of the ionic nature of h-BN causing B and N atoms to respond oppositely to the applied electric field. 
For (i) we use a model consisting of 360C atoms (9x10 orthorhombic unit cells) and recompute the energy curve with the same amount of atoms for graphene, while for (ii) we keep the usual 6x12 orthorhombic cells.
The resulting barriers are shown in Fig.~\ref{curv}.
While the barriers indeed decrease when considering permeation from the side favourable by curvature, the net effect is small compared to the total barrier, $\sim$0.2~eV in the case of a buckled h-BN sheet and less than 0.1~eV in the case of the CNT. These modifications therefore remain insufficient to explain the discrepancies between theory and experiment despite the relatively high curvature induced in these model system. 
Moreover, if it is indeed the case that the process is slow, we should rather expect it to chemisorb as discussed before and therefore only increase the barrier further. 

\section{Hydrogenation}
\begin{figure} \centering\small \hfill
\begin{subfigure}{0.20\textwidth} \centering \includegraphics[height=26mm]{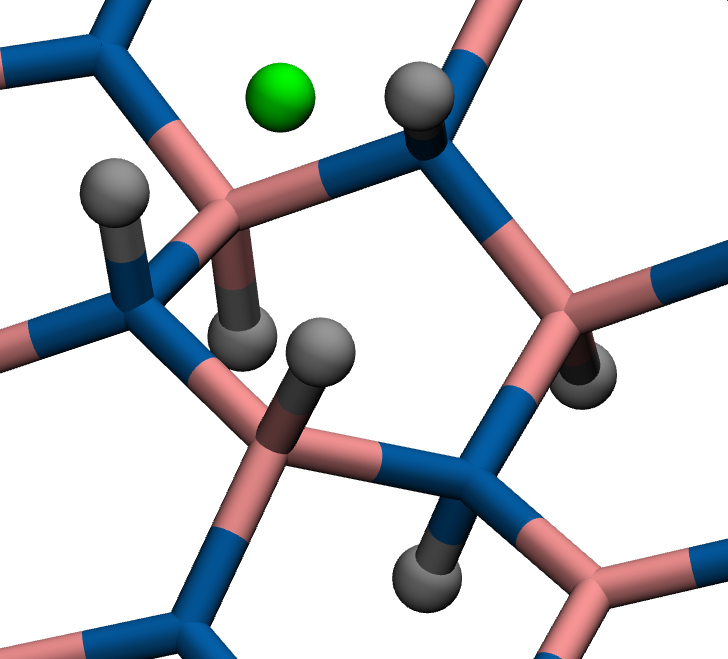} \caption{\centering initial structure}\end{subfigure} \hfill 
\begin{subfigure}{0.20\textwidth} \centering \includegraphics[height=26mm]{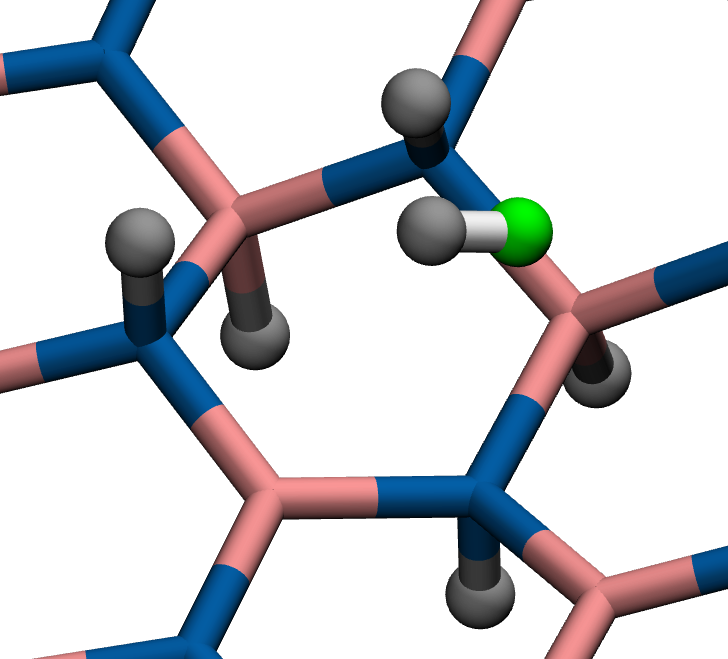} \caption{\centering after optimization}\end{subfigure} \hfill
\hfill
\caption{\label{fig:h2}
  Proton permeation of hydrogenated h-BN: (a) shows the initial state with a proton 3~\AA\ above the h-BN surface with six chemisorbed H atoms and (b) shows the optimized structure with H$_2$ formed.
}
\end{figure}

In Ref.~\cite{kroes_energetics_2016} we considered in detail the hydrogenation of h-BN. 
We also considered the permeation of a proton through these hydrogenated surfaces.
However, as in the case of static permeation in vacuum, we find the initial state with a proton 3~\AA\ above (measured from the center of mass) a group of chemisorbed hydrogen atoms to be unstable.
In this case however, the intermediate state is not a chemisorbed one, but rather the formation of molecular hydrogen (H$_2$) as shown in Fig.~\ref{fig:h2}.
We find similar formations of molecular hydrogen in other structures considered in Ref.~\cite{kroes_energetics_2016}.
As a result, these intermediate stable states again lead to a increase in barrier when optimization of the path is considered.

\section{Conclusion}
We considered several effects on calculated barrier heights for proton permeation through h-BN and graphene. 
Despite our efforts to include more features of the environment we systematically find an increase of the permeation energy barriers rather than the desired decrease needed to match experimental observations. 
Several complications are discussed such as the high polarizability of the membranes and the existence of stable intermediates.
These naturally arising complications have not been considered in previous work and we therefore found it necessary to comment explicitly on them in this work, despite not providing a definitive explanation for the proton permeation barrier heights observed experimentally.
We therefore believe this problem may lie outside the scope of our DFT-PBE approach if the experimental membranes are indeed assumed to be defect-free and the process is slow compared to atomistic timescales.

The research leading to these results received funding from the Foundation for Fundamental Research on Matter (FOM), part of the Netherlands Organisation for Scientific Research (NWO). The work was carried out on the Dutch national e-infrastructure with the support of SURF Cooperative. 
This project has received funding from the European Union's Horizon 2020 research and innovation programme under grant agreement No. 696656 -- GrapheneCore1.
We thank A.K. Geim, W. Brandeburgo and E.J. Meijer for useful discussions. 
\bibliography{refs}
\bibliographystyle{rsc}

\end{document}